\documentclass[structabstract]{aa}  
\usepackage{graphicx}
\usepackage{hyperref}
\usepackage{color}

\usepackage{txfonts}
\begin{document}
   \title{The smoothness of the interstellar extinction curve in the UV}

   \subtitle{Comparison with recent laboratory measurements of PAH mixtures}

   \author{M. Steglich\inst{1}
          \and
          Y. Carpentier\inst{1}\fnmsep\thanks{Present address: Laboratoire de Physique des Lasers, Atomes et Mol\'ecules (PhLAM), UMR CNRS 8523, Centre d'Etudes et de Recherches Lasers et Applications, Universit\'e de Lille 1, 59655 Villeneuve d'Ascq Cedex, France}
          \and
          C. J\"ager\inst{1}
          \and
          F. Huisken\inst{1}
          \and
          H.-J. R\"ader\inst{2}
          \and
          Th. Henning\inst{3}
          }

   \institute{Laboratory Astrophysics Group of the Max Planck Institute for Astronomy at the Friedrich Schiller University Jena,\\
                   Institute of Solid State Physics, Helmholtzweg 3, D-07743 Jena, Germany\\
              \email{m.steglich@web.de, yvain.carpentier@univ-lille1.fr}
         \and
             Max Planck Institute for Polymer Research, Ackermannweg 10, D-55128 Mainz, Germany
         \and
             Max Planck Institute for Astronomy, K\"onigstuhl 17, D-69117 Heidelberg, Germany
             }

   \date{Published in A\&A 540 (2012) A110; received 9 December 2011; accepted 5 March 2012}

% \abstract{}{}{}{}{} 
% 5 {} token are mandatory
 
  \abstract
  % context heading (optional)
  % {} leave it empty if necessary  
   {As revealed by high-resolution spectral investigations in the wavelength range between 300 and 400 nm, the interstellar extinction curve does not display any of the sharp electronic absorption bands that are characteristic for large polyatomic molecules, such as polycyclic aromatic hydrocarbons (PAHs), which belong to the most abundant interstellar molecules.}
  % aims heading (mandatory)
   {We aim to verify whether the absorption curves of mixtures of medium-sized PAHs produced in the laboratory can explain the astronomical observations.}
  % methods heading (mandatory)
   {The PAH mixtures were synthesized by infrared laser pyrolysis and subsequent chemical extraction and size separation. The matrix isolation technique was used to study the absorption spectra of isolated molecules at low temperature.}
  % results heading (mandatory)
   {Our experimental results demonstrate that the UV-visible absorption curves of PAH mixtures can be very smooth, displaying no sharp bands, if the molecular diversity is sufficiently high.}
  % conclusions heading (optional), leave it empty if necessary 
   {In view of the absence of sharp electronic features on the interstellar extinction curve for 300 $< \lambda <$ 400 nm, we conclude from our experimental findings that the interstellar PAH population must be very diverse. The low fractional abundances of individual species prevent their detection on the basis of spectral fingerprints in the UV.}


   \maketitle
%
%________________________________________________________________

\section{Introduction}

Very many absorption features known as ``diffuse interstellar bands'' (DIBs) appear on the interstellar extinction curve for wavelengths longer than 400 nm. The upper wavelength limit below which DIBs have been observed was recently shifted to 1800 nm (Geballe et al. \cite{geballe11}). The origin of the DIBs remains enigmatic, but numerous atomic, molecular, and solid state carriers have been proposed (for a review, see Sarre \cite{sarre06}). Among these proposals, neutral and ionized polycyclic aromatic hydrocarbons (PAHs) are considered to belong to the most promising candidates (Cox \cite{cox11}).

In addition to electronic absorption bands in the visible and near-IR, which are characteristic for PAH ions and certain sufficiently-sized neutral PAHs, these molecules are characterized by strong absorptions in the UV. Recently, astronomers tried to detect the spectral signatures of individual PAHs in the UV part ($\lambda <$ 400 nm) of the interstellar extinction curve (Clayton et al. \cite{clayton03}, Gredel et al. \cite{gredel11}, Salama et al. \cite{salama11}). While Clayton et al. \cite{clayton03} used low-resolution observations in the 157 $-$ 318 nm range, which were obtained with the Space Telescope Imaging Spectrograph of the Hubble Space Telescope, the surveys reported by Gredel et al. \cite{gredel11} and Salama et al. \cite{salama11} covered high-resolution spectra for wavelengths longer than 305 nm recorded with the UVES spectrograph of the Very Large Telescope. However, besides the well-known UV bump at 217.5 nm, some atomic lines, and bands from small, mostly diatomic molecules, such as CH, CH$^+$, CN, OH$^+$, or NH, no narrow bands related to large gas-phase molecules were found. Based on the signal-to-noise ratio in these observations ($S/N >$ 100), very low upper fractional abundances for \textit{specific} PAHs in the diffuse interstellar medium (ISM) on the order of $N$(PAH)/$N$(H) $\approx$ $10^{-10}$ $-$ $10^{-8}$ (Gredel et al. \cite{gredel11}) were derived. On the other hand, inferred from the aromatic infrared emission bands, PAHs are expected to occur in large quantities in the ISM (estimated \textit{total} PAH abundances on the order of $10^{-7}$; Habart et al. \cite{habart04}; Tielens \cite{tielens08}).

Here, we show that mixtures of isolated and cold PAHs, which were produced in the laboratory under astrophysically relevant conditions, can display absorption curves that are smooth in the UV-visible spectral region if the molecular variety is sufficiently high. Hence, our results demonstrate that the astronomical detection of PAHs as a class of interstellar molecules in the infrared does not necessarily contradict the observed UV properties of interstellar extinction.
%__________________________________________________________________

\section{Studied PAH mixtures}

The samples studied here were produced by infrared multiphoton dissociation of a gaseous hydrocarbon precursor (ethene, C$_2$H$_4$) followed by the gas phase synthesis of large carbonaceous molecules and grains. The applied CO$_2$ laser pyrolysis method is described elsewhere (J\"ager et al. \cite{jaeger09}; Steglich et al. \cite{steglich10}). The molecular components of the condensate were extracted with the solvents methanol (CH$_3$OH) or dichloromethane (CH$_2$Cl$_2$). The constituents of these extracts, which could unambiguously be identified, are known PAHs. However, PAH-like molecules with multiple hydrogen saturation, i.e., containing CH$_2$ groups and, thus, displaying aliphatic character, must be present as well (see below and J\"ager et al. \cite{jaeger09}). For simplicity, we will refer to all molecular components as ``PAHs''.

\begin{figure}
   \centering
   \includegraphics[scale=0.32]{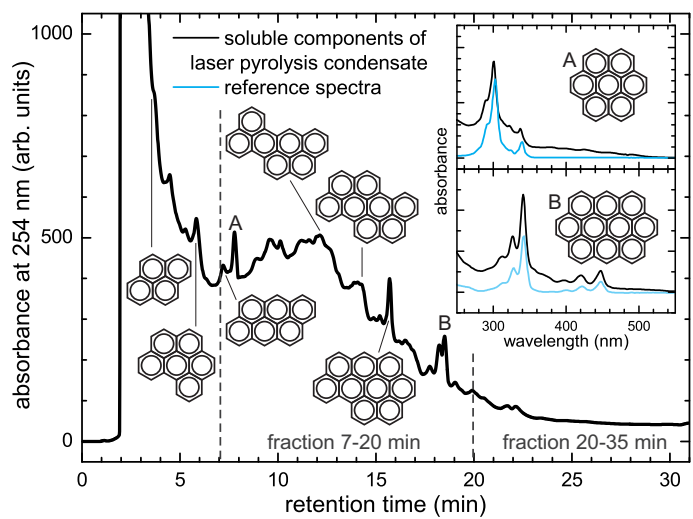}
   \caption{High-performance liquid chromatography spectrum of the soluble components of the laser pyrolysis condensate. The PAHs coronene and ovalene are labeled ``A'' and ``B'', respectively. Inset: Comparison between the UV-visible absorption spectra of coronene and ovalene as measured during the HPLC of the sample (black curves) and their respective reference spectra (light blue curves).}\label{figHPLC}
    \end{figure}

The analysis of the soluble condensate was mainly based on high-performance liquid chromatography (HPLC). With this method, different molecular components in a solvent can be separated according to their interaction time with the stationary phase of a column. For this study, we used a column from Jasco, specifically designed for the analytical fractionation and analysis of PAHs. A size separation of a molecular mixture can be realized since larger PAHs have longer retention times on the column than smaller ones. The chromatographic separation was realized with a two-solvent gradient program running at a temperature of 303 K and using methanol and dichloromethane at a flow rate of 1 ml/min. In combination with the retention time, the characteristic electronic absorption spectrum of each species can be used for identification. In our system, the threshold for the detection of a single species is about 5 ng. The HPLC spectrum of the soluble condensate and the absorption spectra of two selected components, which were identified to be the PAHs coronene (C$_{24}$H$_{12}$) and ovalene (C$_{32}$H$_{14}$), are displayed in Fig. \ref{figHPLC}. In the HPLC spectrum, distinct peaks are created only by the species whose product of ``abundance in the solvent'' and ``absorption strength at a given wavelength'' (here 254 nm) is high. Many other PAHs, hidden in the broad underlying continuum, are difficult to identify.

Since we are mainly interested in the larger PAHs of the condensate, we chose to divide the soluble extract into three different fractions (see the dashed lines in Fig. \ref{figHPLC}). The fraction with retention times between 0$-$7 min contains small molecules up to the size of molecules containing approximately 22 C atoms such as benzo[ghi]perylene (C$_{22}$H$_{12}$). The fraction ``7$-$20 min'' roughly covers the molecular range from anthanthrene (C$_{22}$H$_{12}$) to ovalene (C$_{32}$H$_{14}$). All larger species should be included in the last fraction ``20$-$35 min''. However, there are a few complications hindering a perfect size separation. First, the retention time is only roughly a monotonic function of the PAH size. The specific shape of each molecule (compact, elongated, or bent structure) does play a significant role, too, and can modify the retention time considerably. Second, larger molecules are more difficult to dissolve. The solubility even changes with the presence or absence of smaller species. This can lead to a ``smearing'' of certain individual components over the whole retention time scale. Especially traces of small PAHs with very high solubilities, such as anthracene (C$_{14}$H$_{10}$), phenanthrene (C$_{14}$H$_{10}$), or pyrene (C$_{16}$H$_{10}$), whose main peaks in the HPLC spectrum appear below 7 min, can also be found in the other two fractions that should normally contain only the larger species (see also the matrix spectra below).

\begin{figure}
   \centering
   \includegraphics[scale=0.32]{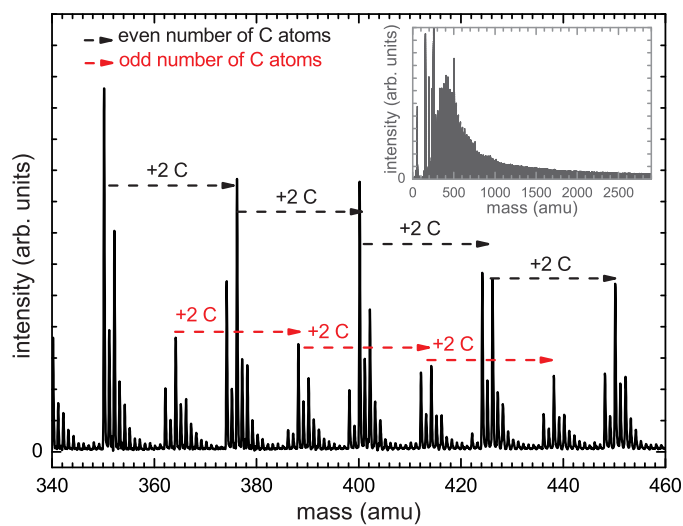}
   \caption{Time-of-flight mass spectrum of the laser pyrolysis condensate. Inset: Overview over wider mass range (taken from J\"ager et al. \cite{jaeger09}).}
              \label{figMALDI}
    \end{figure}

With the HPLC technique, the identification of certain components of the soluble condensate is restricted to well-known molecules whose absorption spectra can be used for comparison. Therefore, only the presence of a few PAHs composed of an even number of C atoms and with ``normal'' H saturation (only aromatic CH groups) could be verified. On the other hand, as revealed by matrix-assisted laser desorption/ionization in combination with time-of-flight mass spectrometry (MALDI-TOF), the condensate is composed of a huge variety of different species. This is demonstrated in Fig. \ref{figMALDI}. In addition to the ``normal'' PAHs with an even number of C atoms, molecules containing an odd number of C atoms are apparent in the mass spectrum. These species, if structurally comparable to the other PAHs, must contain at least one non-aromatic CH$_2$ group to have a closed electronic shell structure. The aliphatic character of some constituents of the laser pyrolysis condensate was also verified via IR absorption spectroscopy, which will be shown in an upcoming publication (for comparison, see also J\"ager et al. \cite{jaeger06}; \cite{jaeger09}). The MALDI-TOF studies have yet another interesting result. Apart from the rather small species that can be analyzed and size-separated by HPLC, very large molecules up to 3000 u, corresponding to PAHs containing more than 200 C atoms, were found (see the inset in Fig. \ref{figMALDI} and J\"ager et al. \cite{jaeger09}). Unfortunately, it is not possible at the moment to enrich mixtures with such large PAHs in sufficient quantities for spectroscopic studies, mainly because of their poor solubility.

\begin{figure}
   \centering
   \includegraphics[scale=0.33]{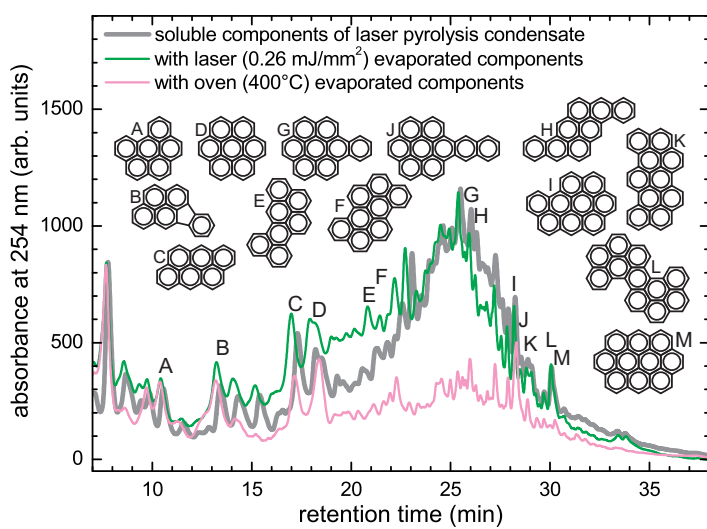}
   \caption{Comparison of the HPLC spectrum of the original condensate obtained by laser pyrolysis with the spectra measured after the same material has been evaporated by either a laser or thermally in an oven. The peaks labeled with the letters ``A'' to ``M'' correspond to the displayed identified PAHs.}\label{figHPLC2}
    \end{figure}

The UV-visible absorption curves of the PAH mixtures presented later were measured by applying the matrix isolation technique. Within this method, the solid extracts have to be transferred to the gas phase before incorporation into the inert gas matrix. For this purpose, we applied thermal evaporation at temperatures $\leq$ 400 $^\circ$C as well as laser desorption with a pulsed Nd:YAG laser, which was operated with 10 Hz at 532 nm ($\leq$ 0.26 mJ/mm$^2$, pulse length ca. 5 ns). To prove how much was actually transferred into the gas phase, the evaporated materials were condensed on a cold shield and subjected to a second HPLC analysis. Compared to the HPLC spectrum of Fig. \ref{figHPLC}, where the HPLC system was optimized for the efficient fractionation of large PAHs, the retention times of the individual components shown in Fig. \ref{figHPLC2} were chosen to be longer, allowing a better tracing of the molecules. Note that the retention times of individual components can vary up to $\approx$ 0.5 min. Comparing the HPLC spectra of the evaporated materials with the spectrum of the original substance, the conclusion can be drawn that the laser desorption method is clearly better suited to transfer the PAH mixtures into the matrix without changing their composition. It follows that matrices doped with laser-evaporated PAHs resemble the original distribution, whereas matrices containing thermally evaporated species are dominated by molecules with high vapor pressures. Therefore, we expect the matrices prepared by laser desorption to reflect the same large variety of different species as encountered in the sample.
%

%__________________________________________________________________

\section{UV-visible absorption spectra of mixtures of matrix-isolated PAHs}

\begin{figure*}
   \centering
   \includegraphics[scale=0.33]{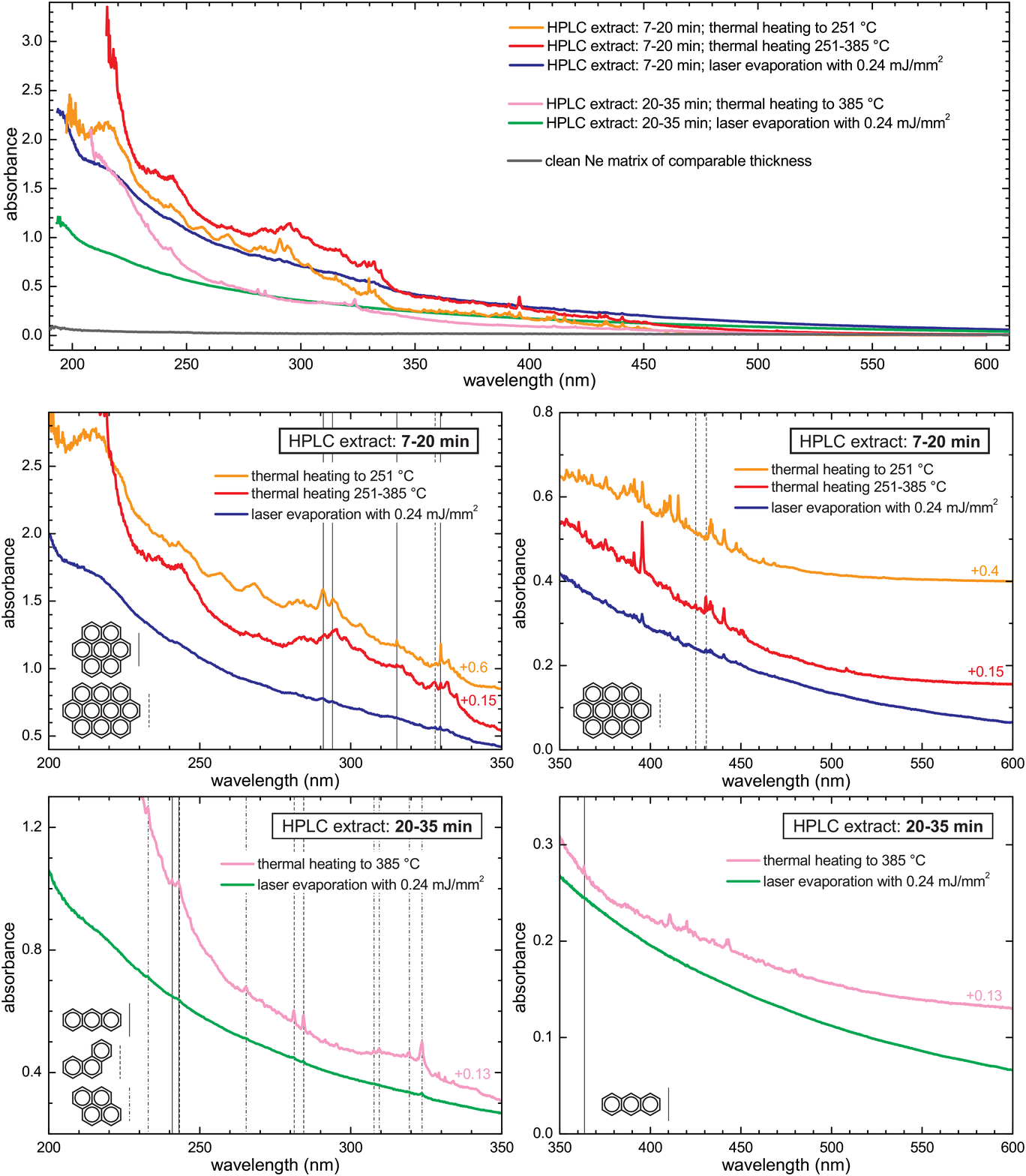}
   \caption{Absorption spectra of PAH mixtures obtained after HPLC fractionization of the soluble laser pyrolysis condensate. The molecules were incorporated into solid Ne matrices kept at 5.5 K. A clean Ne matrix of comparable thickness is shown for comparison. Top panel: Overview over all measurements. The spectra were corrected for weak baseline variations, mainly caused by differences in the sample reflectivity, by shifting them to zero absorbance at 850 nm. Other panels: Numbers given at the right-hand side indicate that the spectra were shifted in $y$-direction for a better comparison. The vertical lines mark the band positions of PAHs that could be identified based on their characteristic peaks in the Ne matrix spectra. The structures of the corresponding molecules are displayed, accordingly.}
              \label{figMIS}
    \end{figure*}

The absorption spectra of the two condensate fractions ``7$-$20 min'' and ``20$-$35 min'' are presented in Fig. \ref{figMIS}. The molecules of the solid extracts were transferred either via laser desorption or by thermal evaporation into the gas phase and, thereafter, deposited with an excess of Ne atoms onto a cold CaF$_2$ window, which was kept at a temperature of 5.5 K. Based on previous experiments with individual PAHs and PAH mixtures, we can exclude the formation of PAH fragments in the vapor phase and their accumulation in the matrix (see also the discussion along with Fig. \ref{figHPLC2}). For comparison, the spectrum of an undoped Ne matrix of similar thickness is displayed in the top panel of Fig. \ref{figMIS}. Under the present experimental conditions, the Ne matrices are supposed to be mainly crystalline (Cradock \& Hinchcliffe \cite{cradock75}). Site effects causing a molecular band to split up in the matrix spectrum can be excluded, i.e, each band in the matrix spectrum has a one-to-one correspondence to the respective band in the gas phase. The undoped matrix does not exhibit any absorption in the investigated wavelength range. The weak contribution to scattering at short wavelengths can almost be neglected. Basically, we also exclude possible additional light scattering in the PAH-doped matrices as the dimensions of the molecules are much smaller than the wavelength. The extinction of all shown spectra is mainly due to PAH absorption.\footnote{For comparison, see also the spectra of individual PAHs in Ne matrices and as solid films (Steglich et al. \cite{steglich10}, \cite{steglich11}), where the light scattering can be neglected compared to the absorption.} Because of small baseline variations, primarily caused by different sample reflectivities, we shifted all measured spectra at 850 nm to zero absorbance.

As already stated above, the matrices that were doped by laser vaporization contain a larger variety of different species. Obviously, this leads to quite smooth UV-visible absorption curves with almost no fine structure for both fractions. Although this continuous absorption caused by close-lying and overlapping bands is present in all spectra, the matrices prepared by thermal evaporation additionally feature many sharp signatures caused by an overabundance of species with high vapor pressure. A few of these sharp bands originate from known PAHs, such as coronene and ovalene, already previously identified in the HPLC extract ``7$-$20 min''. Unfortunately, many features have to remain unidentified because appropriate spectra of individual PAHs that are larger than, or equal in size to, anthanthrene, isolated in solid Ne are not available for comparison. At least some of the peaks of unknown origin in the matrix spectra probably belong to PAHs that could already be verified via HPLC (see Figs. \ref{figHPLC} and \ref{figHPLC2}).

The matrix spectra of the fraction ``20$-$35 min'', which should contain larger molecules composed of more than 32 C atoms, display only a few weak and narrow bands. Except for the bands at wavelengths longer than 400 nm, they can be considered to be impurities because they belong to the very small, high vapor pressure PAHs anthracene, phenanthrene, and pyrene. As previously mentioned, trace amounts of these small molecules are present in all fractions as a result of imperfect size separation by HPLC. Finally, note that, while the matrices prepared by thermal evaporation exhibit stronger absorbance at short wavelengths compared to the corresponding matrices prepared by laser desorption, the situation is reversed at long wavelengths. The spectra intersect each other somewhere between 300 and 400 nm (see the top panel in Fig. \ref{figMIS}). This can be easily explained by differences in the fractional abundances of larger PAHs with stronger contribution to absorption at longer wavelengths. 

To summarize, if the original variety of larger PAHs in the size distribution is kept, as is the case when laser desorption is used, the resulting Ne matrix spectrum reveals a fairly smooth decay for wavelengths longer than $\sim$ 250 nm. We do not observe any sharp absorption band that can be assigned to a specific PAH. On the other hand, based on the measured absorbance, we can be sure that the larger PAHs were indeed incorporated into the matrix.
%__________________________________________________________________

\section{Astrophysical implications -- extension to larger species in the gas phase}

The experiments described above demonstrate that a PAH mixture composed of a sufficient amount of different species exhibits a featureless absorption curve. However, there are a few points that have to be considered if we aim to apply our experimental results to PAH mixtures in the ISM. The first and most obvious point concerns the perturbations induced by the weak interaction forces between the studied molecules and the Ne atoms of the host matrix. Compared to the spectra of completely free gas-phase PAHs, the absorption bands of the individual species in the matrix are redshifted and broadened. Obviously, the broadening promotes the creation of a smooth absorption curve, and some narrow features could perhaps be visible if the PAHs were isolated in the gas phase. However, the matrix-induced broadening usually affects only those bands that involve the first one or two electronic transitions in a closed-shell molecule, i.e., S$_0$$\rightarrow$S$_{1,2}$. The bands of these transitions can be very narrow in the gas phase and would appear substantially broadened in a Ne matrix (see, e.g., Gredel et al. \cite{gredel11}). Nevertheless, as inferred from the aromatic infrared emission bands, the bulk of interstellar PAHs are probably larger than the species we studied here ($\gtrsim$ 40 C atoms; Tielens \cite{tielens08}). The first electronic transitions of these larger molecules are generally located at wavelengths longer than $\approx$ 400 nm (see, e.g., Ruiterkamp et al. \cite{ruiterkamp02}; Malloci et al. \cite{malloci07}). At shorter wavelengths, the bandwidths are usually governed by the short lifetimes of the excited (high-energy) states and by vibrational interactions with energetically lower-lying states (Rouill\'e et al. \cite{rouille09}), i.e., they should be very similar for completely free and matrix-isolated molecules (Steglich et al. \cite{steglich11}). Therefore, a smooth extinction curve at $\lambda \lesssim$ 400 nm can be expected if the molecular variety of the interstellar PAH population is at least comparable to our laboratory-produced mixtures. Probably, the interstellar PAHs show an even higher diversity because the number of isomers raises rapidly with the size of the molecules. Furthermore, free-flying PAHs are exposed in the interstellar space to ionizing irradiation. The formation of ions, which, compared to their neutral precursors, display even broader bands below 400 nm (Steglich et al. \cite{steglich11}), additionally increases the molecular diversity.

\begin{figure}
   \centering
   \includegraphics[scale=0.31]{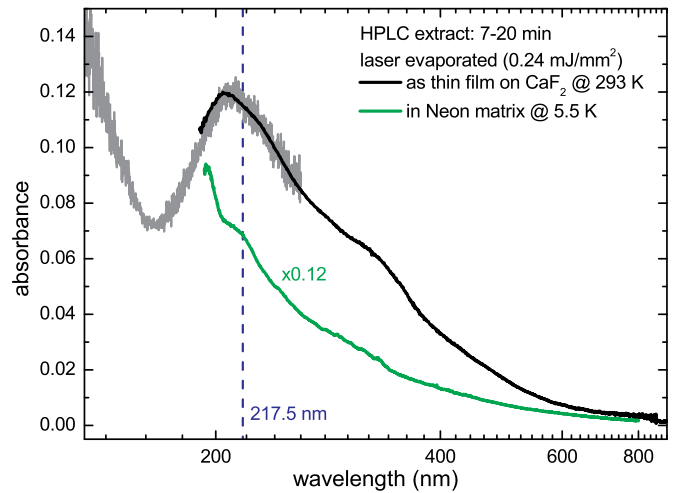}
   \caption{Absorption spectrum of a film-like deposit of the HPLC fraction ``7$-$20 min'' compared with the spectrum of the same sample in the Ne matrix. The more noisy gray part of the spectrum was recorded with a vacuum-UV spectrometer.}
              \label{figBUMP}
    \end{figure}

A non-negligible part of the interstellar PAHs might also be locked in clusters or on the surface of grains (see, e.g., Fig. 6 in Tielens \cite{tielens08}). The absorption spectrum of larger clusters, if bonded only by van der Waals forces, can be simulated in the laboratory by measuring the spectrum of a thin film of PAHs deposited on a transparent substrate. An example is presented in Fig. \ref{figBUMP}. The film-like deposit of the extract ``7$-$20 min'' was prepared by laser desorption and subsequent deposition onto a 5.5 K cold CaF$_2$ window (without simultaneous Ne flow). The individual molecules of the film were allowed to relax into energetically favored positions by keeping the sample at room temperature and under vacuum for several hours.\footnote{Upon cooling the sample back to cryogenic temperatures afterward, the absorption spectrum does not change anymore. The widths and positions of the spectral features are mainly governed by the structural arrangement of the molecules.} The resulting absorption spectrum does not display any narrow feature. Only a broad bump is apparent at 208 nm. For comparison, the Ne matrix spectrum of the laser-evaporated extract ``7$-$20 min'' is also included in Fig. \ref{figBUMP}. There, the bump appears considerably narrower and further to the blue at 195 nm. It is expected to shift to longer wavelengths for mixtures with a larger mean size of the aromatic planes, approaching the position of the interstellar UV bump at 217.5 nm (Steglich et al. \cite{steglich10}).
%______________________________________________________________

\section{Conclusions}
We presented experimental UV-visible absorption spectra of mixtures of cold and isolated PAHs, which were produced in the laboratory under astrophysically relevant conditions. We showed that the spectra can be almost completely featureless if the molecular variety is sufficiently high. The present results and their interpretation are also in line with another recent study implying that the DIBs, unlike the interstellar UV bump, very probably do not originate from PAHs (Steglich et al. \cite{steglich11}). This is furthermore supported by an investigation from Xiang et al. (\cite{xiang11}), who were unable to find any correlation between the 2175 $\AA$ feature and nine of the strongest DIBs. Finally, if we assume that the optical emission bands in the Red Rectangle Nebula are caused by some of the DIB carriers (see Sarre \cite{sarre06}), there is yet another argument against a connection between the interstellar PAH population and the DIBs because no correlation was found between these optical bands and the aromatic infrared emission at 3.3 $\mu$m (Sarre \cite{sarre06}).

In summary, we propose that similar to the aromatic emission bands in the infrared, the interstellar PAHs give only rise to collective features also in the UV-visible, i.e., an enhanced but rather smooth extinction culminating in a broad bump at 2175 $\AA$, and not individual fingerprints, which could be used for an identification of specific molecules. The absence of narrow absorption bands related to larger gas-phase polyaromatic molecules on the interstellar extinction curve for 300 nm $< \lambda <$ 400 nm can be explained by a high molecular diversity of the interstellar PAH population. The fractional abundances of individual species are too low to allow their detection on the basis of their characteristic electronic spectra. Nevertheless, high-resolution observations at wavelengths shorter than 300 nm are still lacking. Such measurements could help to verify whether or not one can detect the electronic fingerprints of individual large organic molecules in the ISM.

\begin{acknowledgements}
Part of this work was supported by the \emph{Deut\-sche For\-schungs\-ge\-mein\-schaft, DFG\/} project number Hu 474/21-2. We thank H. Mutschke for providing access to the vacuum-UV spectrometer and G. Born for technical assistance with the chemical extraction and HPLC operation.
\end{acknowledgements}
%______________________________________________________________

\end{document}